\documentclass[aps,twocolumn,preprintnumbers,amsmath,amssymb,superscriptaddress,nofootinbib,english]{revtex4-1}
\pdfoutput=1
\usepackage{times,amsmath,amsfonts,amssymb}
\usepackage{graphicx}
\usepackage{dcolumn}
\usepackage{bm}
\usepackage{enumerate}
\usepackage{epsfig}
\usepackage{hyperref}
\usepackage[usenames]{color}
\usepackage{url}
\usepackage[normalem]{ulem}
\usepackage[T1]{fontenc}
\usepackage{amsmath}
\usepackage{longtable}
\usepackage{changepage}

\def\sc{\scriptscriptstyle}
\def\be{\begin{equation}}
\def\ee{\end{equation}}
\def\ben{\begin{eqnarray}}
\def\een{\end{eqnarray}}
\def\ba{\begin{array}}
\def\ea{\end{array}}

\newcommand{\bq}{\begin{eqnarray}}
\newcommand{\eq}{\end{eqnarray}}
\newcommand{\bes}{\begin{subequations}}
\newcommand{\ees}{\end{subequations}}

\definecolor{orange}{rgb}{1,0.3,0}
\definecolor{vert}{rgb}{0.1,0.5,0.25}

\begin{document}
\newcommand{\half}{{\textstyle\frac{1}{2}}}
\allowdisplaybreaks[3]
\def\triangledown{\nabla}
\def\grad3{\hat{\nabla}}
\def\a{\alpha}
\def\b{\beta}
\def\g{\gamma}\def\G{\Gamma}
\def\d{\delta}\def\D{\Delta}
\def\ep{\epsilon}
\def\et{\eta}
\def\z{\zeta}
\def\t{\theta}\def\T{\Theta}
\def\l{\lambda}\def\L{\Lambda}
\def\m{\mu}
\def\f{\phi}\def\F{\Phi}
\def\n{\nu}
\def\r{\rho}
\def\s{\sigma}\def\S{\Sigma}
\def\ta{\tau}
\def\x{\chi}
\def\o{\omega}\def\O{\Omega}
\def\k{\kappa}
\def\pa {\partial}
\def\ov{\over}
\def\br{\\}
\def\ud{\underline}

\def\lcdm{\Lambda{\rm CDM}}
\def\qcdm{{\rm QCDM}}
\def\nloc{RR}
\def\msun{M_{\odot}/h}
\def\dw{f(\Box^{-1}R)}
\def\costhe{{\rm cos}\theta}
\def\sinthe{{\rm sin}\theta}
\def\cosphi{{\rm cos}\varphi}
\def\sinphi{{\rm sin}\varphi}
\def\sintwothe{{\rm sin}^2\theta}
\def\costwothe{{\rm cos}^2\theta}
\def\sintwophi{{\rm sin}^2\varphi}
\def\costwophi{{\rm cos}^2\varphi}
\def\boxa{L = 250\ {\rm Mpc}/h}
\def\boxb{L = 100\ {\rm Mpc}/h}
\def\boxc{L = 400\ {\rm Mpc}/h}
\def\ttwo{h_c \leq 0.03\ \left[{\rm Mpc}/h\right]}
\def\tfour{h_c \leq 0.12\ \left[{\rm Mpc}/h\right]}
\def\tfive{h_c \leq 0.24\ \left[{\rm Mpc}/h\right]}

\def\lplanck{\Lambda{\rm CDM}^{Planck}}
\def\nplanck{RR^{Planck}}
\def\ndirian{RR^{Dirian}}
\def\nhubl{RR^{Planck, \neq H_0}}
\def\njune{RR^{Planck, \neq \Sigma m_\nu}}
\def\nneut{RR^{Planck, \neq \Sigma m_\nu, H_0}}

\def\pxpx{\left[\partial_x\partial_x\Phi\right]}
\def\pypy{\left[\partial_y\partial_y\Phi\right]}
\def\pzpz{\left[\partial_z\partial_z\Phi\right]}
\def\pxpy{\left[\partial_x\partial_y\Phi\right]}
\def\pxpz{\left[\partial_x\partial_z\Phi\right]}
\def\pypz{\left[\partial_y\partial_z\Phi\right]}

\newcommand{\ramses}{{\sc ramses}}
\newcommand{\isis}{{\sc{isis}}}
\newcommand{\isisns}{{\sc{isis-nonstatic}}}
\newcommand{\ecosmog}{{\sc ecosmog}}
\newcommand{\dgpm}{{\sc dgpm}}

\title{Changing the Bayesian prior: Absolute neutrino mass constraints in nonlocal gravity}

\thanks{Based on observations obtained with Planck (http://www.esa.int/Planck), an ESA science mission with instruments and contributions directly funded by ESA Member States, NASA, and Canada.}

\author{Yves Dirian}
\email[Electronic address: ]{yves.dirian@unige.ch}
\affiliation{D\'{e}partement de Physique Th\'{e}orique and Center for Astroparticle Physics,
Universit\'{e} de Gen\`{e}ve, 24 quai Ansermet, CH-1211 Gen\`{e}ve 4, Switzerland}

\begin{abstract}
Prior change is discussed in observational constraints studies of nonlocally modified gravity, where a model characterized by a modification of the form $\sim m^2 R\Box^{-2}R$ to the Einstein-Hilbert action was compared against the base $\lcdm$ one in a Bayesian way. It was found that the competing modified gravity model is significantly disfavored (at $22 \,$:$\, 1$ in terms of betting-odds) against $\lcdm$ given CMB+SNIa+BAO data, because of a tension appearing in the $H_0 \,$--$\, \Omega_M$ plane. We identify the underlying mechanism generating such a tension and show that it is mostly caused by the late-time, quite smooth, phantom nature of the effective dark energy described by the nonlocal model. We find that the tension is resolved by considering an extension of the initial baseline, consisting in allowing the absolute mass of three degenerated massive neutrino species $\sum m_\nu /3$ to take values within a prior interval consistent with existing data. As a net effect, the absolute neutrino mass is inferred to be non-vanishing at $2 \sigma$ level, best-fitting at $\sum m_\nu \approx 0.21 {\, \rm eV}$, and the Bayesian tension disappears rendering the nonlocal gravity model statistically equivalent to $\lcdm$, given recent CMB+SNIa+BAO data. We also discuss constraints from growth rate measurements $f \sigma_8$, whose fit is found to be improved by a larger massive neutrino fraction as well. The $\nu$-extended nonlocal model also prefers a higher value of $H_0$ than $\lcdm$, therefore in better agreement with local measurements. Our study provides one more example suggesting that the neutrino density fraction $\Omega_{\nu}$ is partially degenerated with the nature of the dark energy. This emphasizes the importance of cosmological and terrestrial neutrino research and, as a massive neutrino background impacts structure formation observables non-negligibly, proves to be especially relevant for future galaxy surveys.
\end{abstract} 

\maketitle

\section{Introduction}\label{sec:intro}

Modern cosmology has undergone fast developments over the past two decades. 
The evidence for accelerated expansion from observations of distant Type Ia supernovae \cite{Riess:1998cb,Perlmutter:1998np} led to the abandon of the inflationary cold dark matter (\textsc{CDM}) paradigm and to the birth of the $\lcdm$ one, introducing a cosmological constant $\Lambda$ into the model.
Such an extension adds an overwhelming, thereby constant, dark energy component to the corresponding homogeneous and isotropic Friedmann-Lema\^itre-Robertson-Walker (FLRW) universe and at the same time raises fundamental theoretical questions about its origin, late-time domination and naturalness \cite{Carroll:2003qq,Bianchi:2010uw}.
The idea of modelling differently such a component for addressing these issues keeps to be actively explored nowadays through different perspectives, in particular in attempting to modify General Relativity (GR) (see e.g. \cite{Nojiri:2010wj, Clifton:2011jh, Joyce:2014kja, Berti:2015itd, Koyama:2015vza} for recent reviews). Among these, nonlocal modifications to GR have attracted much interest over the years. They are motivated by radiative corrections induced by gravity or light-matter components on curved spacetime \cite{1987NuPhB.282..163B, BARVINSKY1990471, Gorbar:2002pw}, also responsible for the trace anomaly \cite{Antoniadis:1991fa, Barvinsky:1994cg} or non-perturbative effects giving raise to a scale dependent Newton's constant \cite{ArkaniHamed:2002fu, Barvinsky:2003kg, Hamber:2005dw, LopezNacir:2006tn, Barvinsky:2011hd}, but also by renormalizability and singularity issues in higher-derivative theories of gravity \cite{2012PhRvL.108c1101B,2012PhRvD..86d4005M} or string inspired scenarios \cite{Biswas:2005qr,2015PhRvD..91l4059C}. Models have also been proposed in a bottom-up approach for appreciating the effects of various types of nonlocal operators within the cosmological context \cite{Wetterich:1997bz, Espriu2005197, Nojiri:2007uq, Deser:2007jk, Koivisto:2008xfa, Deser:2013uya, Maggiore:2013mea, Ferreira:2013tqn, Maggiore:2014sia, Mitsou:2015yfa, Cusin:2015rex, Cusin:2016nzi, Nersisyan:2016jta, Vardanyan:2017kal}.

Two nonlocal models of this sort have been recently proposed, one of which is characterized by the addition of a term $\sim m^2 \big( g_{\mu \nu} \Box^{-1} R \big)^T$ to Einstein's equations \cite{Maggiore:2013mea}, whereas a second one modifies the Einstein-Hilbert action by a term $\sim m^2 R \Box^{-2} R$ \cite{Maggiore:2014sia}. Their effective dark energy phenomenology has been studied in Ref. \cite{Dirian:2014ara} where it was found that both models describe a quite smooth (i.e. whose perturbations are small), phantom dark energy component emerging at late-times, with an equation-of-state today of $w_{\rm DE}=-1.04$ and $w_{\rm DE}=-1.15$ respectively. Furthermore, the models feature a fifth force that enhances the clustering of linear structures compared to that in $\lcdm$: at $\lesssim 6 \%$ level in the linear matter power spectrum around the BAO scale (see also Ref. \cite{2014JCAP...09..031B} where nonlinear structure formation through N-body simulation has been studied for the model of Ref. \cite{Maggiore:2014sia}). Observational constraints using Bayesian techniques were then carried out in Refs. \cite{Dirian:2014bma,Dirian:2016puz} in a complementary perspective \cite{Tegmark:1998ab, Eisenstein:1998hr}, i.e. using cosmological data from Cosmic Microwave Background (CMB), Baryon Acoustic Oscillations (BAO) and Type Ia supernova (SNIa) observations. Constraints on the linear growth rate of structures were also obtained a posteriori, that is, constraining the quantity $f \sigma_8$ derived from both nonlocal models on their respective CMB+SNIa+BAO bestfit with Redshift-Space Distortions (RSD) measurements. The joined CMB+SNIa+BAO constraints showed that, provided a prior parametrization fixed on the so-called \textit{Planck} 2015 baseline \cite{Ade:2015xua}, the model of Ref. \cite{Maggiore:2013mea} is indistinguishable from standard $\lcdm$ with a Bayes factor of $1.0$, whereas the one presented in Ref. \cite{Maggiore:2014sia} is significantly disfavored with a Bayes factor of $22.7$. The latter discrepancy was shown to result from a severe CMB-SNIa tension appearing in the $H_0 \,$--$\, \Omega_M$ plane, breaking the nonlocal models' concordance given the data.

In this article, we analyze in more details the origin of this tension and find a solution for resolving it. In particular, we will see that changing the neutrino sector of the aforementioned baseline from one dominant active mass-eigenstate to three degenerated ones, whose absolute mass is taken as a free parameter, restores the concordance of the nonlocal model to a non-negligible extent. Effectively, such a resolution exploits degeneracies between modified gravity effects of the nonlocal model and those caused by a more massive neutrino component. Similar degeneracies have already been noticed in local modified gravity theories, for instance at linear level in TeVeS \cite{PhysRevLett.96.011301}, covariant galileons \cite{2014PhRvD..90b3528B}, $K$-mouflage \cite{2015PhRvD..91f3528B} and recently in Horndeski models \cite{Bellomo:2016xhl}, but also at the nonlinear one through N-body simulations of $f(R)$ scenarios in Ref. \cite{2014MNRAS.440...75B}.   

The rest of this paper is organized as follows. The specific nonlocal gravity model we consider will briefly be reviewed in section \ref{sec:model} . In section \ref{sec:cosmo_data} , we expose how the latter is embedded into a statistical model, given specific data that we also present. Furthermore, we specify our prior parametrization and justify such a choice from empirical evidences. Section \ref{sec:res} presents the results drawn from observational constraints, attempts for a comprehensive analysis of the resolution of the aforementioned tension and quantifies it using Bayesian model comparison methods. We present a summary of our work in section \ref{sec:conc} before we conclude.

\section{The $\nloc$ nonlocal gravity model}\label{sec:model}

In this section, we briefly introduce the nonlocal gravity model originally proposed by Ref.~\cite{Maggiore:2014sia} which, following Ref.~\cite{Dirian:2016puz} for conveniency, will be referred to as the $\nloc$ model. Our conventions, notations and strategy to solve the equations are similar to that of Refs.~\cite{Dirian:2016puz,Dirian:2014ara}, that the interested reader is invited to consult for more details.

\medskip

The $\nloc$ model is defined through an extension of the Einstein-Hilbert action reading,
\bq\label{eq:action}
S_{\nloc} = \frac{1}{16\pi G}\int {\rm d}^4 x\sqrt{-g}\left[R - \frac{m^2}{6}R\Box^{-2}R - \mathcal{L}_m\right],
\eq
where $\mathcal{L}_m$ is the Lagrange density of minimally coupled matter fields and $\Box^{-1}$ is a formal notation for a Green's function of the curved-space d'Alembert operator $\Box \equiv \nabla^\mu\nabla_\mu$. Since the kernel of $\Box$ is non-trivial such a Green's function is not unique and its precise structure in Eq.~\eqref{eq:action} is not known, as well as the origin of the mass scale $m$. This originates from the fact that this model was built through a bottom-up approach and its embedding into a more fundamental framework still remains under investigation \cite{Maggiore:2015rma, Maggiore:2016fbn}, see also Ref. \cite{Maggiore:2016gpx} for a recent review. From a more technical point of view, several subtleties arise from the presence of $\Box^{-1}$ into an action. One of them concerns the causal character of the classical evolution governed by the corresponding Euler-Lagrange equations and has been discussed in details in the literature (see e.g. Refs. \cite{Soussa:2003vv, Barvinsky:2003kg}). Another one ties to the ``localized'' version of the action \eqref{eq:action}, most conveniently used for computing the corresponding cosmology without having to deal with less manipulable integro-differential systems. The present work makes use of such a local version, taken to be\footnote{See also Ref. \cite{2014JCAP...10..065D} where another localization has been proposed.}
\bq\label{eq:action-local}
S_{\nloc, \mathrm{loc}}~ &&= \frac{1}{16\pi G}\int {\rm d}^4 x\sqrt{-g}\left[R - \frac{m^2}{6}R V - \xi_1\left(\Box U + R\right) \right. \nonumber \\
&&\ \ \ \ \ \ \ \ \ \ \ \ \  \ \ \ \ \ \ \ \ \ \ \ \ \ \ \ \ \ \ \ \left. - \xi_2\left(\Box V + U\right) - \mathcal{L}_m\right],
\eq
where $U$ and $V$ are two ``auxiliary'' scalar fields and $\xi_{1,2}$ are Lagrange multipliers enforcing the constraints,
\bq
\label{eq:u}\Box U &=& - R , \ \ \ \ \ \ \\
\label{eq:s}\Box V &=& - U  .
\eq
sourcing the latter. Invoking a given (left) inverse, one can solve the latter formally in simply writing,
\bq
\label{eq:iu} U &=& - \Box^{-1} R \, , \ \ \ \ \ \ \\
\label{eq:is} V &=& - \Box^{-1} U = \Box^{-2} R \, ,
\eq
which allows one to integrate $U$, $V$ out from the action (and $\xi_{1,2}$ in a similar way), leading back to the nonlocal action \eqref{eq:action}. The same procedure can be applied at the level of the equations of motion, that are
\bq
&&\label{eq:fe1}G_{\mu\nu} - \frac{m^2}{6}K_{\mu\nu} = 8 \pi G \, T_{\mu\nu} \, , \\
&&\label{eq:fe2}\Box U = -R \, , \\
&&\label{eq:fe3}\Box V = -U \, ,
\eq
with
\bq
\label{eq:fe4}K_{\mu\nu} \equiv 2 V G_{\mu\nu} - 2\nabla_\mu\nabla_\nu V - 2\nabla_{(\mu}U\nabla_{\nu)}V \nonumber \\ 
+ \left(2\Box V + \nabla_\alpha U\nabla^\alpha V - \frac{U^2}{2}\right)g_{\mu\nu} \, ,
\eq
where $T^{\mu\nu} \equiv \left(2/\sqrt{-g}\right)\delta\left(\mathcal{L}_m\sqrt{-g}\right)/\delta g_{\mu\nu}$ is the matter source.
In writing Eqs.~\eqref{eq:iu} and \eqref{eq:is} at the level of the equations of motion, causality of the classical dynamics requires that the used source-convolving kernels are all of the retarded kind and, in accordance with their artificial nature, that the auxiliary fields possess no homogeneous (i.e. free) solutions \cite{Koshelev:2008ie, Koivisto:2009jn, Barvinsky:2011rk, Deser:2013uya, Maggiore:2013mea, Foffa:2013sma, Foffa:2013vma, Maggiore:2014sia, 2014JCAP...10..065D}, at least into the \textit{in} state. 
This uniquely fixes the inverse of $\Box$ used in integrating out $U$ and $V$ and catches the classical solutions that we are interested in in that work. At the ``localized'' level, trivial homogeneous modes for the auxiliary fields is realized in fixing vanishing values for $U$ and $V$ and their time derivatives on the initial hypersurface. Here we interpret these requirements as theory-level data that one should supplement to the local action \eqref{eq:action-local} from the beginning, and changing these prescriptions changes the underlying nonlocal model.

According to the procedure used until now \cite{Dirian:2014ara, 2015JCAP...04..044D, Dirian:2016puz}, we set the initial hypersurface deep into the radiation dominated era (RD). During this period, at the background level, the curvature scalar is sub-dominating compared to the overall energy scale of the process, so one can naively set the space-averaged value $\left. \bar{R} \right|_{\mathrm{RD}}\simeq 0$ and see that the resulting $\left. \bar{U},\bar{V}  \right|_{\mathrm{RD}}$ are therefore not sourced, and remain small. Linear perturbations $\left. \delta U, \delta V  \right|_{\mathrm{RD}}$ induced by gravitational ones remain also small \cite{2014JCAP...10..065D}. However, once matter starts to dominate over radiation, the latter quantities acquire a non-trivial dynamics leading to the emergence of a late time, quite smooth dynamical dark energy component \cite{Dirian:2014ara} driving the accelerated expansion of the Universe which ends up into a Big Rip \cite{2014JCAP...10..065D,Caldwell:2003vq}.
For being ultimately legitimate, this choice of initial conditions needs to assume that the secular growth of $U$ and $V$ during earlier stages of the Universe is mild enough for not strongly affecting the vanishing values chosen while starting in RD. For the case of inflationary scenarios, it can be argued that the additional term in Eq.~\eqref{eq:action} is suppressed by the corresponding inflationary energy scale, since it only modifies the theory into the far-infrared. In any case this question deserves a special attention and a more quantitative analysis is needed (see Refs. \cite{Maggiore:2015rma,Cusin:2016nzi,Codello:2016xhm} where such a scenario was considered). Of course, the configuration reached by the auxiliary fields in RD will depend on the particular assumed model of inflation but also on other processes affecting the scalar curvature within the primordial Universe, such as the electroweak or QCD phase transition, conformal anomalies, or during RD itself through the presence of thermalized Standard Model massive particles \cite{Caldwell:2013mox}. This issue was recently anticipated in Ref. \cite{Nersisyan:2016hjh}, where the authors studied the effect of varying the auxiliary fields' initial conditions deep into RD over a broad range of values, unveiling in particular the existence of another phenomenologically viable cosmology of the nonlocal model \eqref{eq:action}. 

We close this description in emphasizing that, from a statistical perspective which will be taken in the following, fixing vanishing initial conditions for the auxiliary fields is a \textit{modelling assumption} or \textit{theoretical prior} (much as pretending that $\Lambda$-dark energy is replaced by the $RR$ one in our context). Therefore, the only remaining free parameter added to the Einstein-Hilbert action is the mass scale $m$, whose value is to be fixed from spatial flatness condition -- similarly of fixing the value of $\Lambda$ in flat $\lcdm$ models. The model has therefore the same number of free parameters as the $\lcdm$ model.

\section{Cosmological models}\label{sec:cosmo_data}

The cosmological models incorporating $\Lambda$- and $\nloc$-dark energies considered in this article are one-parameter extensions of those studied in Ref.~\cite{Dirian:2016puz} and are similarly tested within a Bayesian framework. \\
In this section, we therefore only provide a brief, self-contained review of the datasets and parametrization used and refer the reader to Ref.~\cite{Dirian:2016puz}, and references therein, for further details about their structure and construction. Technical details relative to the equations and numerical implementation used can be found in Sec. 2 and App. A. of Ref.~\cite{Dirian:2016puz}. 

\subsection{Datasets}

For performing our global fit we utilize Cosmic Microwave Background (CMB) data complemented with distant Type Ia supernovae (SNIa) observations and distance measurements from several Baryon Acoustic Oscillations (BAO) surveys. Together with CMB lensing data, SNIa and BAO observations allow to apply constraints on the late time expansion of the Universe since they break further primary CMB degeneracies. We will then discuss a posteriori constraints on the growth rate from Redshift-Space Distortions (RSD) measurements.

The CMB dataset we consider comes from the \textit{Planck} satellite mission and is made of the {\it lowTEB} power spectra data for multipoles $\ell \leq 29$, the high-$\ell$ TT, TE and EE ones for $\ell > 29$ \cite{Aghanim:2015xee}, as well as the data relative to the power spectrum of the reconstructed CMB lensing potential \cite{Ade:2015zua}. In the following, this dataset will be referred to as the \textit{Planck} dataset. For the BAO data, we include the isotropic measurements reported in Ref.~\cite{2011MNRAS.416.3017B} at $z_{\rm eff} = 0.106$ from 6dfGS and Ref.~\cite{2015MNRAS.449..835R} at $z_{\rm eff} = 0.15$ from SDSS-MGS DR7, as well as the anisotropic ones of Ref.~\cite{2014MNRAS.441...24A} from the LOWZ ($z_{\rm eff} = 0.32$)  and CMASS ($z_{\rm eff} = 0.57$) samples from the BOSS DR11 release. The SNIa data we use are those of the SDSS-II/SNLS combined analysis (also known as \textit{JLA}), which comprises $740$ objects at $z \lesssim 1$ \cite{2014A&A...568A..22B}.

Below we study constraints with two dataset combinations: (i) \textit{Planck} and (ii) the combined \textit{Planck}, BAO and SNIa data, which we shall refer to as the \textit{BAPJ} dataset to preserve the nomenclature of Ref.~\cite{Dirian:2016puz}.

\subsection{Parametrization and prior specification}

In accordance with the use of the above measurements, the complement of the dark energy sector of the models considered in this paper are defined following the so-called \textit{Planck {\rm 2015} baseline} \cite{Ade:2015xua}. 
Regarding the radiation/matter content of the Universe, ionization history and primordial initial conditions, the baseline assumes a particular modelling that has now become the standard cosmological setting on which is built the \textit{base} $\lcdm$ model. Predictions from such a model are computed in using a cosmological linear Einstein-Boltzmann code such as \textsc{CAMB} \cite{Lewis:1999bs} or \textsc{CLASS} \cite{2011JCAP...07..034B}. In our study, we make use of the latter and provide a modified version of it on \textsc{GitHub} (see \cite{git_nonlocal} for the link) which includes the $RR$ gravity model introduced in the previous subsection. Bayesian parameter extraction and model selection are carried out with the Markov Chain Monte Carlo (MCMC) code \textsc{MONTEPYTHON} \cite{2013JCAP...02..001A} originally interfaced with \textsc{CLASS}.

The baseline specifies a continuous 6-dimensional parametrization which can be provided by the vector,
\bq 
\theta_{\rm base} = \big( H_0, 100 \, \omega_b, \omega_{cdm}, {\rm ln} (10^{10}A_s ), n_s, \tau \big) , \label{base_param}
\eq
with $H_0 \equiv 100 h ~ {\rm km/s/Mpc}$, the Hubble expansion rate today; $\omega_i \equiv \Omega_i h^2$, where $\Omega_i$ is the present energy density fraction of baryons ($i=b$) and cold dark matter ($i=cdm$); $A_s$ and $n_s$ are the amplitude and tilt of the power spectrum of primordial fluctuations respectively and $\tau$ is the optical depth to reionization. In the following, we will also use derived quantities as $\Omega_m$ the total matter density fraction today, $\sigma_8$ the root mean square linear matter fluctuations in a sphere of radius $8 {\rm \, Mpc/h}$ at $z=0$ and the dark energy density fractions $\Omega_\Lambda$, $\Omega_{\nloc}$ corresponding to the $\lcdm$ and $\nloc$ models respectively. The latter are determined in requiring a vanishing spatial curvature of the Universe $\Omega_K=0$ and this is done adequately tuning the parameter controlling the dark energy density within the model specified: $\Lambda$ in $\lcdm$ or $m$ in $\nloc$. In this work we use improper flat prior with edges everywhere unbounded expect for the lower bound of the optical depth $\tau$ taken to be $0.01$, in accordance with Gunn-Peterson trough observations (see e.g. Ref. \cite{Becker:2001ee}).

In the neutrino sector, the baseline assumes one dominant active neutrino species of fixed mass $M_\nu = 0.06 {\, \rm eV}$, i.e. with prior
\bq 
P \big( M_\nu | \lcdm) = \delta \big( M_\nu / {\rm eV} - 0.06 \big) \, ,
\eq
whose temperature is tuned for reproducing predictions from neutrino decoupling computations \cite{Mangano:2005cc}. The massive neutrino species is supplemented with a massless neutrino one whose effective number of degrees-of-freedom is adjusted so as to reproduce a total effective number of relativistic components of $N_{\rm eff}=3.046$ in the early Universe (see e.g. Refs. \cite{Archidiacono:2013fha, 2014NJPh...16f5002L}). 

As will be justified more clearly below, the extension we consider in this work assumes three active massive species with degenerated mass-eigenstates, i.e. the sum of their masses respects $\sum m_\nu \equiv 3 m_0$, where the absolute neutrino mass $m_0$ is taken to be a free parameter. Their temperature and the additional massless neutrino component are fixed respecting the same conditions as those of the initial baseline. 
We therefore adopt the following parametrization,
\bq
\theta_{\rm \nu} = \big( H_0, 100 \, \omega_b, \omega_{cdm}, {\rm ln} (10^{10}A_s ), n_s, \tau, \Sigma \, m_\nu \big) \, ,~~~ \label{nu_param}
\eq
and the prior interval on the additional parameter is taken to be uniform and compact $\sum  m_\nu \in [ 0, 5 ] {\, \rm eV}$, consistent with existing data (see subsection \ref{subsec:nu} for more details and references on the latter choice).

The cosmological models denoted by $\lcdm$ and $\nloc$ are parametrized by $\theta_{\rm base}$ while their extended versions build out of $\theta_{\rm \nu}$ will be called $\nu \lcdm$ and $\nu \nloc$ respectively. 

\begin{table*}
\begin{adjustwidth}{-0cm}{}
\caption{Summary of the means, standard deviations and (effective) $\chi^2$ goodness-of-fit values for the one-dimensional marginalized likelihood distributions of the $\lcdm$, $\nu \lcdm$, $RR$ and $\nu RR$ models obtained with the \textit{Planck} dataset. The $\Delta\chi^2$ values are taken with respect to the $\lcdm$ $\chi^2$ values for each dataset, where $\chi^2 \equiv -2 \, {\rm ln} \mathcal{L}$, with $\mathcal{L}$ being the likelihood function. All bounds shown correspond to $1\sigma$ unless explicitly stated otherwise.}
\centering
\begin{tabular}{@{}ccccccccccc}
\hline\hline
  & \ \ $Planck$ \ \ & $Planck$ \ \ &$Planck$ \ \ &$Planck$ \\
  & \ \ $\lcdm$ \ \ & $\nu\lcdm$ & \ \ $RR$ \ \ & $\nu RR$  \\
\hline 
\\
$100\omega_b$  &\ \ $2.225^{+0.016}_{-0.016}$ & \ \ $2.220^{+0.017}_{-0.017}$ &\ \ $2.227^{+0.016}_{-0.016}$ &\ \ $2.222^{+0.017}_{-0.017}$\\
$\omega_{cdm}$  &\ \ $0.1194^{+0.0014}_{-0.0015}$ &\ \ $0.1198^{+0.0015}_{-0.0016}$ &\ \ $0.1191^{+0.0014}_{-0.0015}$ &\ \ $0.1196^{+0.0015}_{-0.0016}$\\ 
$H_0$  &\ \ $67.50^{+0.65}_{-0.66}$ &\ \ $66.12^{+2.1}_{-1.2}$ &\ \ $71.51^{+0.81}_{-0.84}$ &\ \ $69.57^{+2.5}_{-1.6}$\\
${\rm ln}\left(10^{10}A_s\right)$  &\ \ $3.064^{+0.025}_{-0.025}$ &\ \ $3.080^{+0.030}_{-0.034}$ &\ \ $3.047^{+0.026}_{-0.025}$ &\ \ $3.071^{+0.032}_{-0.035}$\\ 
$n_s$  &\ \ $0.9647^{+0.0048}_{-0.0049}$ &\ \ $0.9637^{+0.0050}_{-0.0050}$ &\ \ $0.9649^{+0.0049}_{-0.0049}$ &\ \ $0.9639^{+0.0051}_{-0.0052}$ \\
$\tau$  &\ \ $0.06530^{+0.014}_{-0.014}$ &\ \ $0.07312^{+0.016}_{-0.018}$ &\ \ $0.05733^{+0.014}_{-0.014}$ &\ \ $0.06905^{+0.017}_{-0.018}$\\
$\sum m_\nu\ [{\rm eV}]$ &\ \ $0.06\ {\rm (fixed)}$ &\ \ $<0.50 \,\, (2 \sigma) \,\,$ &\ \ $0.06\ {\rm (fixed)}$ &\ \ $<0.51 \,\, (2 \sigma)$\\
\\
$\sigma_8$  &\ \ $0.8171^{+0.0089}_{-0.0089}$ &\ \ $0.7949^{+0.033}_{-0.016}$ &\ \ $0.8487^{+0.0097}_{-0.0096}$ &\ \ $0.8212^{+0.038}_{-0.020}$  & \\
\\
$\Delta\chi^2_{Planck}$   &\ \ $0\ (\chi^2 = 12943.30)$ &\ \ $-0.04$ &\ \ $-1.6$ &\ \ $-1.6$ \\  
\\
\hline \hline
\end{tabular}
\label{table:planck}
\end{adjustwidth}
\end{table*}

\begin{table*}
\begin{adjustwidth}{-0cm}{}
\caption{As for Table I but obtained with the \textit{BAPJ} dataset.}
\centering
\begin{tabular}{@{}ccccccccccc}
\hline \hline
  & \ \ $BAPJ$ \ \ &\ \  $BAPJ$ \ \ &\ \  $BAPJ$ \ \ &\ \  $BAPJ$  \\
  & \ \  $\lcdm$ \ \ & $\nu\lcdm$ & \ \ $RR$ \ \ & $\nu RR$    \\
\hline 
\\
$100\omega_b$  & \ \ $2.228^{+0.014}_{-0.015}$ & \ \ $2.229^{+0.014}_{-0.015}$ &\ \ $2.213^{+0.014}_{-0.015}$ &\ \ $2.221^{+0.014}_{-0.015}$ & \\
$\omega_{cdm}$  &\ \ $0.1190^{+0.0011}_{-0.0011}$ &\ \ $0.1189^{+0.0011}_{-0.0011}$ &\ \ $0.1210^{+0.0010}_{-0.0010}$ &\ \ $0.1197^{+0.0012}_{-0.0012}$ \\ 
$H_0$ &\ \ $67.67^{+0.47}_{-0.50}$ &\ \ $67.60^{+0.66}_{-0.55}$ &\ \ $70.44^{+0.56}_{-0.56}$ &\ \ $69.49^{+0.79}_{-0.80}$ \\
${\rm ln}\left(10^{10}A_s\right)$  &\ \ $3.066^{+0.019}_{-0.026}$ &\ \ $3.071^{+0.026}_{-0.029}$ &\ \ $3.027^{+0.027}_{-0.023}$ &\ \ $3.071^{+0.032}_{-0.032}$ \\ 
$n_s$  &\ \ $0.9656^{+0.0041}_{-0.0043}$ &\ \ $0.9661^{+0.0043}_{-0.0043}$ &\ \ $0.9601^{+0.0040}_{-0.0039}$ &\ \ $0.9635^{+0.0043}_{-0.0045}$ \\
$\tau$  &\ \ $0.06678^{+0.011}_{-0.013}$ &\ \ $0.06965^{+0.014}_{-0.015}$ &\ \ $0.04516^{+0.014}_{-0.012}$ &\ \ $0.06880^{+0.017}_{-0.017}$ \\
$\sum m_\nu\ [{\rm eV}]$ &\ \ $0.06\ {\rm (fixed)}$ &\ \ $< 0.21 \,\, (2 \sigma)$ &\ \ $0.06\ {\rm (fixed)}$ &\ \ $0.219^{+0.083}_{-0.084}$\\
\\
$\sigma_8$  &\ \ $0.8170^{+0.0076}_{-0.0095}$ &\ \ $0.8157^{+0.013}_{-0.011}$ &\ \ $0.8443^{+0.010}_{-0.0099}$ &\ \ $0.8215^{+0.017}_{-0.017}$ \\
\\
$\Delta\chi^2_{Planck}$ &\ \ $0\ (\chi^2 = 12943.42)$ &\ \ $-0.14$ &\ \ $-0.14$ &\ \ $-1.52$ \\  
$\Delta\chi^2_{BAO}$    &\ \ $0\ (\chi^2 = 4.42)$ &\ \ $0$ &\ \ $2.48$ &\ \ $2.38$ \\
$\Delta\chi^2_{JLA}$     &\ \ $0\ (\chi^2 = 683.2)$ &\ \ $-0.12$ &\ \ $3.56$ &\ \ $2.5$ \\
\\
$\Delta\chi^2_{total}$   &\ \  $0\ (\chi^2 = 13631.04)$ &\ \ $-0.26$ &\ \ $5.9$ &\ \ $3.36$ \\
\\
\hline
\hline
\end{tabular}
\label{table:bapj}
\end{adjustwidth}
\end{table*}

\section{Observational constraints and model comparison}\label{sec:res}

In this section, we present and study our observational constraints and model comparison results. \\
We start by analyzing some structural features of the $\nloc$ model and, comparing them to the ones of standard $\lcdm$, we explain the origin of the tensions first found in Ref.~\cite{2015JCAP...04..044D}, given the nonlocal model. Then, we present the parameter constraints on the $\nu RR$ model introduced above and discuss Bayesian model selection through the computation of various Bayes factors, comparing pairs of the four above-mentioned models. We finally also discuss the capabilities of growth rate measurements to constrain further the cosmologies at hand.

\subsection{The origin of the CMB-SNIa tension given $\nloc$}

\begin{figure}
	\centering
	\includegraphics[scale=0.44]{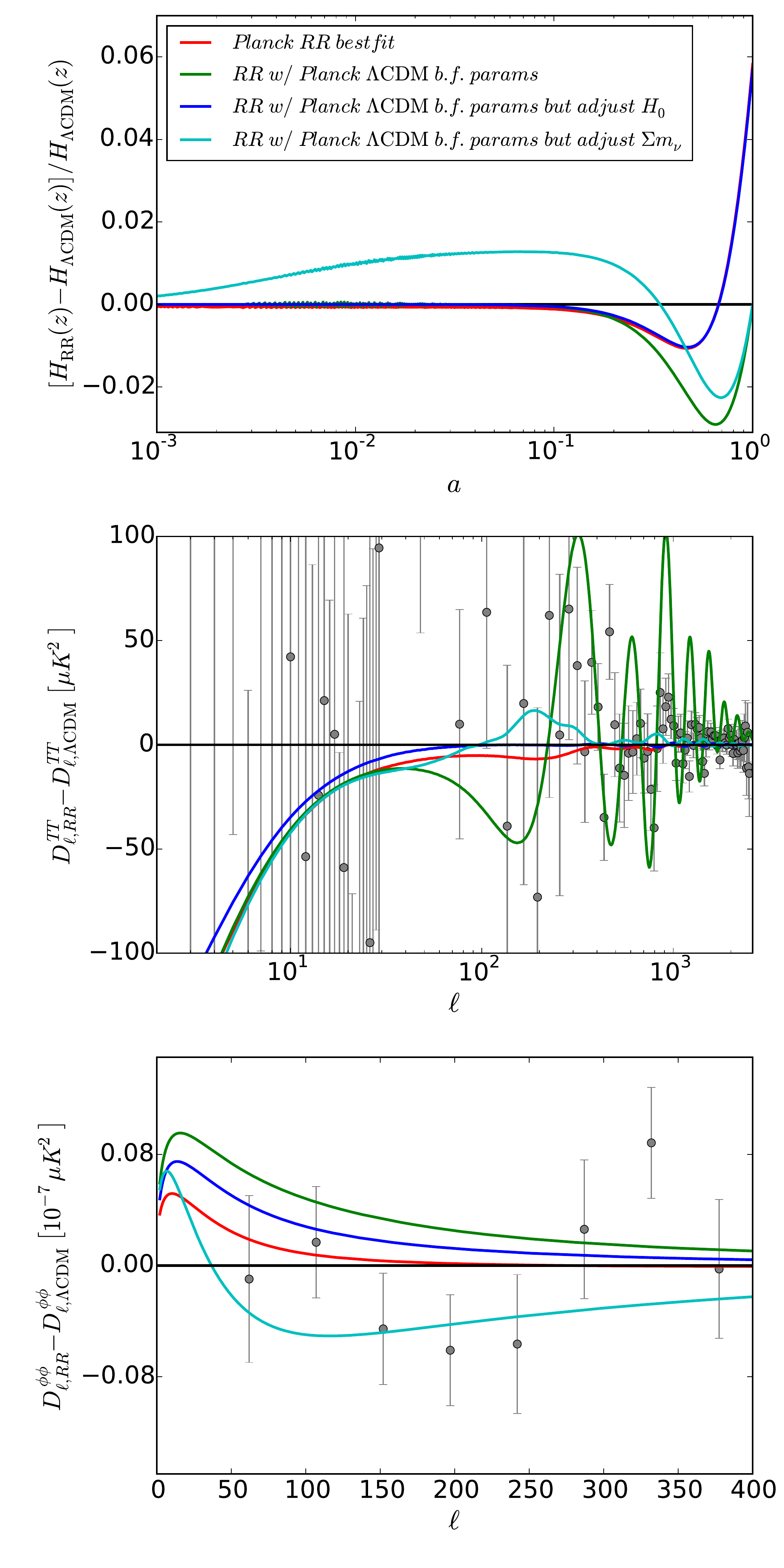}
	\caption{Hubble expansion rate (top), CMB temperature power spectrum (middle) and CMB lensing power spectrum (bottom) for a few illustrative $RR$ cosmologies, plotted as the relative difference to the best-fitting $\lcdm$ cosmology to the \textit{Planck} dataset. The red curve displays the prediction of the best-fitting $RR$ model to the \textit{Planck} dataset. The green curves show the prediction of the $RR$ gravity model with the same parameters as the best-fitting $\lcdm$ model to \textit{Planck} data. The remainder curves show the same as the green ones, but with $H_0 = 71.31 {\rm km/s/Mpc}$ (blue) and $\sum m_\nu = 0.423\ {\rm eV}$ (cyan), which have been adjusted to yield the same angular acoustic scale $\theta_* = 0.010414$ as $\lcdm$. In the middle and lower panels, the grey symbols with errorbars show the power spectra as measured by the \textit{Planck} satellite \cite{Ade:2015xua}.}
\label{fig:h}
\end{figure}

\subsubsection{$Planck$ dataset}

Table \ref{table:planck} summarizes the constraints on the $\lcdm$ and $RR$ models obtained with the \textit{Planck} dataset (second and fourth columns). The parameter shifts between the two models can be understood in comparing the relevant features in both models with each other.

A first noteworthy point is that the differences between the results in the two cosmologies are statistically non-significant ($\lesssim 1 \sigma$) for all parameters in $\theta_{\rm base}$, with the exception of the background-related parameter $H_0$ which undergoes the most significant shift ($\sim 5 \sigma$). Notice that $\sigma_8$, derived at the linear level, also undergoes a significant change ($\sim 3 \sigma$). As already noticed in Ref. \cite{2014JCAP...09..031B}, the latter results from the enhanced clustering in the nonlocal model, mostly induced by a lower expansion rate (see top panel of Fig.~\ref{fig:h}) that reduces the Hubble friction to matter perturbations. This effect is supplemented by, although to a milder level, a higher late-time gravitational strength modelled by a time-dependent effective Newton constant $G_{\rm eff}(z,k)$, exhibiting scale-dependence in the far-infrared \cite{Dirian:2016puz}. More clustering also increases the lensing power (see bottom panel of Fig.~\ref{fig:h}) and in turn smooths out temperature fluctuations more efficiently. Moreover, this requires a smaller primordial amplitude $A_s$ which comes together with a delayed reionization epoch, given that the CMB damping tail constrains well the combination $A_s \, e^{-2 \tau}$ at high-$\ell$ \footnote{Observe that the preference for a lower optical depth to reionization $\tau$ within the nonlocal model compared to $\lcdm$ is consistent with the results found from the new analysis of the \textit{Planck} HFI data \cite{Aghanim:2016yuo}.}. 

As already mentioned above, the \textit{Planck} dataset constrains $H_0$ to be larger in $RR$ ($H_0 \approx 71.51 \pm 0.84\ {\rm km/s/Mpc}$) than in $\lcdm$ ($H_0 \approx 67.50 \pm 0.66\ {\rm km/s/Mpc}$). This preference for higher $H_0$ a fortiori originates from the late-time emerging, quite smooth and phantom nature of the $\nloc$ effective dark energy compared to that modelled by a cosmological constant $\Lambda$. In the following we attempt to provide a comprehensive explanation of this fact. 
The Friedmann equation including a dynamical dark energy component with equation of state $w_{\rm DE}(z)$ reads
\bq 
H(z) = H_0 \big[ \Omega(z) + \Omega_{\rm DE}(z) \big]^{1/2} \, , \label{eq:Friedmann}
\eq
where $\Omega_{\rm DE}(z)$ denotes the dark energy density fraction present in the Universe at redshift $z$ and $\Omega (z)$ includes all the other components, that is, the density fraction of cold dark matter $\Omega_{cdm}(z)$ and of baryons $\Omega_{b}(z)$ in the case of the baseline, but also other ingredients available in extensions of it such as the density fraction of massive neutrinos $\Omega_\nu$. From the dark energy conservation equation, $\Omega_{\rm DE}(z)$ can be written in terms of the dark energy equation of state,
\bq
\Omega_{\rm DE}(z) &=& \Omega_{\rm DE}  ~ \exp \bigg( 3 \int^z_0 {\rm d} z' \frac{1+w_{\rm DE}(z')}{(1+z')} \bigg) \, ,
\eq
which, at low redshift when the dark energy is dominant, can be approximated by
\bq
\Omega_{\rm DE}(z \approx 0) & \simeq & \Omega_{\rm DE} ~ \big(1+3 z \, \delta w_0 \big) \, , \label{eq:Om_de_approx}
\eq
where we wrote $w_{\rm DE}(z \approx 0) \simeq -1+\delta w_0$, with $| \delta w_0 | \ll 1$ and constant, which is sensible enough for the present discussion. One can see that for the case of a phantom dark energy, such as the one featured by the $\nloc$ model for which $\delta w_0 < 0$, the predicted dark energy density $\Omega_{RR}(z)$ is generically smaller than $\Omega_\Lambda$ at $z \gtrsim 0$. This explains why $H_{RR}(z)$ is lower compared to $H_{\Lambda}(z)$, fixing the other parameter values. This is illustrated by the green line in the upper panel of Fig.~\ref{fig:h}, which shows the time evolution of the Hubble rate in the $RR$ model when assuming the cosmological parameters of the best-fitting $\lcdm$ model to the \textit{Planck} dataset. Cosmological constraints on the $RR$ model that are sensitive to $H(z)$ will then generically infer primarily higher values of $H_0$, $\Omega_{cdm}$, $\Omega_b$, etc, for trying to compensate the change induced by the different dark energy modelling. Such a change clearly depends on the particular data set used to derive the constraints. We explore the case of the \textit{Planck} data into this subsection whereas we examine the SNIa \textit{JLA} ones into the next one.

From the point of view of \textit{Planck} data, modifications to $H(z)$ at low redshift alter the predicted angular acoustic scale $\theta_*$, which determines the position of the acoustic peaks of the CMB temperature power spectrum. $\theta_*$ is measured with a very good precision by \textit{Planck} ($\lesssim 0.1 \%$ at $1 \sigma$ in the case of base $\lcdm$) and is robust under cosmology change. It is expressed as $\theta_* \equiv r_*/D_A(z_*)$, where $r_*$ is the sound horizon at the redshift of recombination $z_*$ and $D_A(z_*)$ the comoving angular diameter distance to recombination defined as,
\bq\label{eq:rs}
r_* &\equiv& \int_{z_*}^{\infty} \frac{c_s}{H(z)}{\rm \, d}z \, , \\
D_A(z_*) &\equiv& \int_0^{z_*} \frac{{\rm d}z}{H(z)} \, ,
\eq
where $c_s$ is the sound speed of the primordial plasma
\bq
c_s = 1/\sqrt{3\left[1 + 3\Omega_b/(4\Omega_\gamma)\right]} \, ,
\eq 
with $\Omega_{\gamma}$ the photon density fraction today.
For fixed cosmological parameter values, $r_*$ does not change significantly from $\lcdm$ to $RR$ because it is a function of early time background configurations and does not depend on the particular dark energy modelling. At late time however, such a modelling becomes important and the lower expansion rate in the $RR$ model leads to a larger $D_A(z_*)$, which in turn lowers $\theta_*$. The lower acoustic scale shifts the CMB temperature power spectrum towards higher multipoles $\ell$, yielding the poor fit to the data seen in the middle panel of Fig.~\ref{fig:h}.

In the case of the \textit{Planck} baseline, this discrepancy can be resolved in shifting either the background quantities $H_0$, $\omega_b$ or $\omega_{cdm}$ (or equivalently $\Omega_b$ or $\Omega_{cdm}$) present into the $H(z)$ expression Eq.~\eqref{eq:Friedmann}. However, the shape information of the first CMB peaks such as their relative position and their relative height provide strong, model-independent constraints on both $\omega_b$ and $\omega_{cdm}$ (see e.g. Refs. \cite{Ade:2013zuv, Aghanim:2016sns}) and there is therefore only significant room for $H_0$ to vary. Consequently, since the dark energy featured by the $\nloc$ model is phantom, $H_0$ is doomed to increase. The blue curves in Fig.~\ref{fig:h} show the same as the green ones, but with $H_0$ adjusted to $H_0 = 71.31\ {\rm km/s/Mpc}$ so as to yield the same $\theta_*$ as in the best-fitting $\lcdm$ model to \textit{Planck}. This yields a cosmological scenario that is very similar to the best-fitting $\nloc$ model to \textit{Planck} (red curves in Fig.~\ref{fig:h}), whose goodness-of-fit to \textit{Planck} is better than the base $\lcdm$ with $\Delta \chi^2 = 1.6$. This is mostly because of the lower power in the low-$\ell$ part of the CMB temperature power spectrum, induced by a smaller ISW effect dominating at large-scales. Such a preference being ``inconclusive'' according to the classification reported in our Table \ref{table:stat}, both models are therefore statistically equivalent given \textit{Planck} 2015 CMB data. 

\subsubsection{\textit{BAPJ} dataset}

\begin{figure}
	\centering
	\includegraphics[scale=0.60]{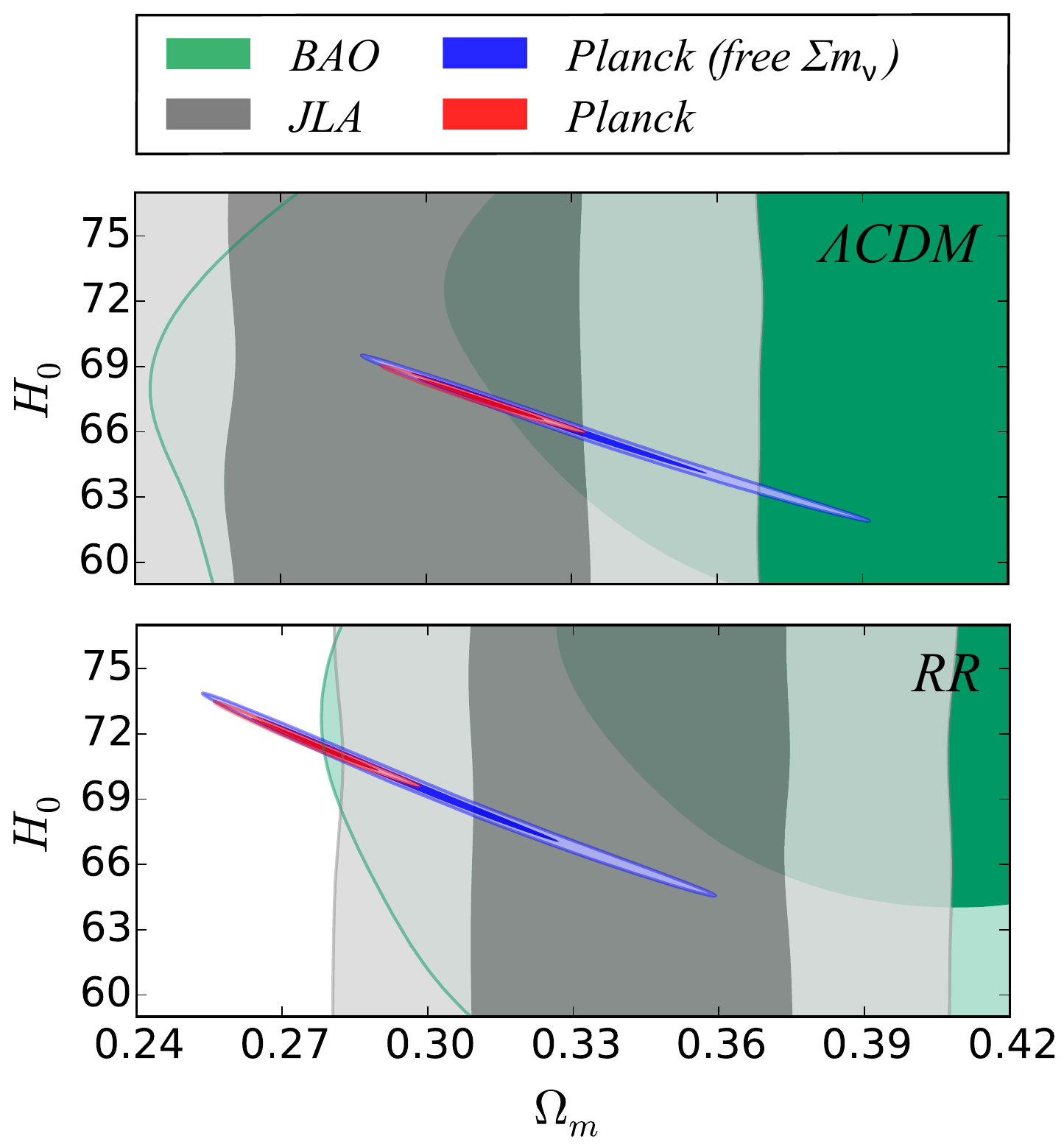}
	\caption{Two dimensional marginalized constraints on the $H_0-\Omega_{m}$ plane in the $\lcdm$ (top) and $RR$ (bottom) models obtained with the \textit{Planck} (red), BAO (green) and JLA (grey) datasets. The blue contours are the same as the red ones, but for constraints in which $\sum m_\nu$ is a free parameter. For fixed color, the two contour shades indicate $1\sigma$ and $2\sigma$ confidence level. The BAO and JLA contours do not change appreciably when $\sum m_\nu$ varies so we do not display them explicitly.}
\label{fig:hom}
\end{figure}

The agreement with observations of the $RR$ model however degrades when it is confronted against the \textit{BAPJ} dataset. The observational tensions that arise when one includes the BAO and SNIa data in the analyzes are better illustrated in Fig.~\ref{fig:hom}. The figure shows the 2d marginalized constraints in the $H_0 \,$--$\, \Omega_{m}$ plane for $\lcdm$ (upper panel) and $RR$ (lower panel), obtained individually using the \textit{Planck} dataset (red), SNIa data (grey) and BAO data (green). Contrary to $\lcdm$, for the $RR$ model the marginalized posterior suggests a $\sim 3 \,$--$\, 4 \sigma$ level tension between \textit{Planck} and SNIa data. According to the above discussion, this can be understood in looking at the luminosity distance relevant for SNIa lightcurves,
\bq
D_L(z) \equiv (1+z) \int^{z}_0 \frac{{\rm d} z'}{H(z')} \, , \label{eq:dL}
\eq
where the expression for $H(z)$ is found in Eq.~\eqref{eq:Friedmann}. SNIa measurements only constrain the total matter density $\Omega_{m}$, whereas $H_0$ has been integrated out via marginalization on the absolute magnitude. Here the fact that the dark energy in the $\nloc$ model is on the phantom side has the net effect of raising $\Omega_{m}$ towards higher values in $\nloc$ than in $\lcdm$, for fixed luminosity distance.  
This shift has already been reported in Ref. \cite{Dirian:2014ara} and is illustrated by the grey contours in Fig.~\ref{fig:hom}. We find $\left. \Omega_{m} \right|_{\lcdm}=0.298 \pm 0.035$ and $\left. \Omega_{m} \right|_{\nloc}=0.343 \pm 0.033$, exhibiting a $\sim 1\sigma$ shift between the two models, given SNIa \textit{JLA} data.

This trend is inconsistent with the one inferred from \textit{Planck}: this explains the origin of the tension quantified at $\Delta \chi^2=5.9$, which appears when constraining the $\nloc$ model using joined \textit{Planck} and SNIa \textit{JLA} data. 

\subsection{Changing the prior: From $\nloc$ to $\nu \nloc$}

\begin{figure*}
	\centering
	\includegraphics[scale=0.85]{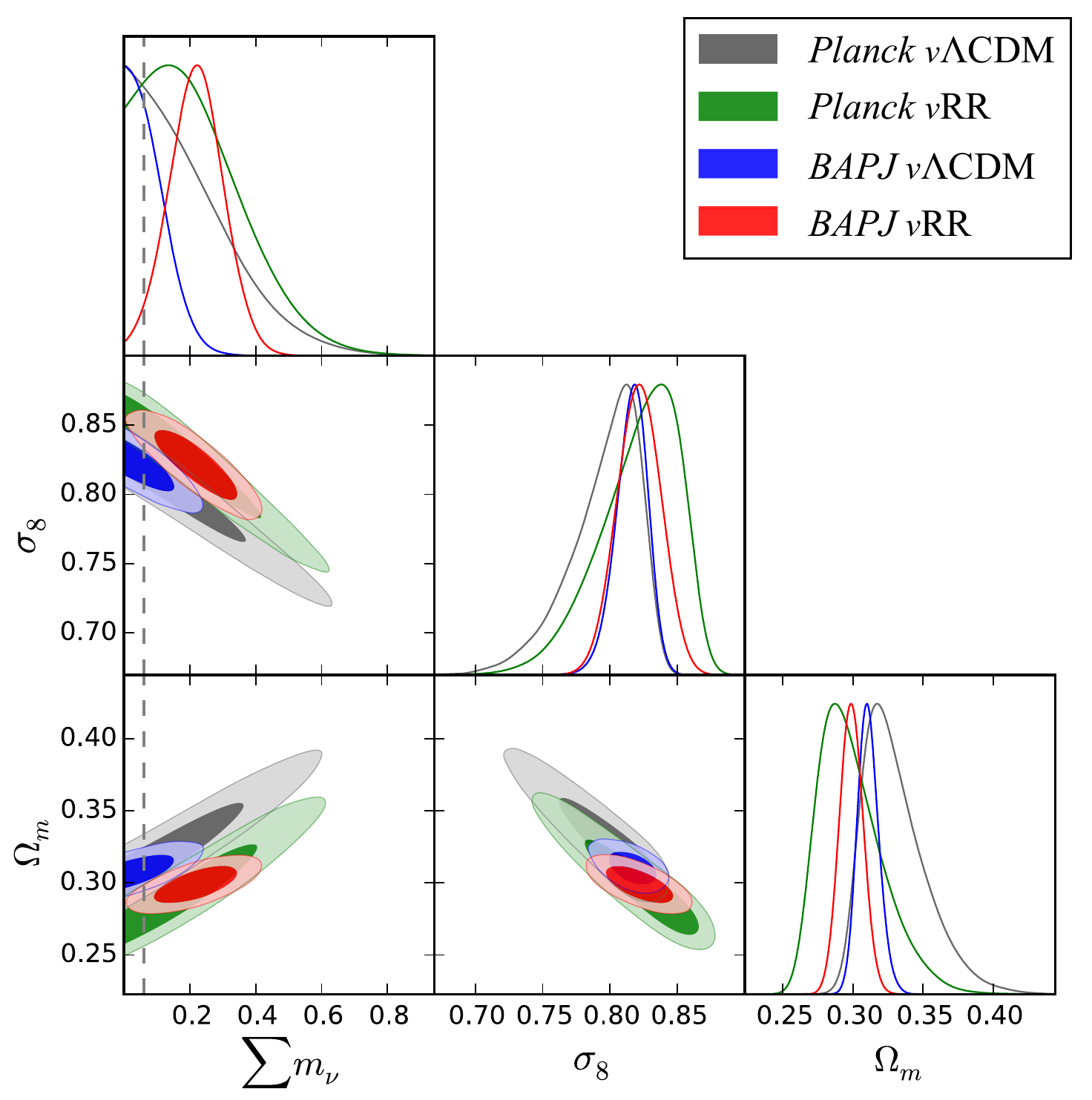}
	\caption{One and two dimensional marginalized constraints on the parameters $\sum m_\nu$, $\sigma_8$ and $\Omega_{m}$ in the $\nu RR$ and $\nu\lcdm$ models, obtained with the \textit{Planck} and \textit{BAPJ} datasets, as labelled. For fixed color, the two contour shades indicate $1\sigma$ and $2\sigma$ limits.}
\label{fig:tri}
\end{figure*}

From the above discussion we have deduced that the late-time phantom nature of the $RR$ effective dark energy induces an increase in $H_0$ given \textit{Planck} data, as this helps to resolve the mismatch with the CMB peaks position constraining $\theta_*$. However, the same fact also induces an increase in $\Omega_{m}$ given SNIa data, which is inconsistent with \textit{Planck}'s preference since the latter provides tight constraints on $\sim \Omega_{m} h^2$ which, together with an increase in $H_0$, forces $\Omega_{m} $ to go down\footnote{This behavior is a generic one for reasonably smooth phantom dark energy, as indicated by the degenerate directions in the $\Omega_{m} \!$ -- $\! w_0$ plane while constraining ${w_0 \rm CDM}$ models given equivalent CMB and SNIa data as those used in this work (see e.g. Fig. 16 of Ref. \cite{2014A&A...568A..22B}).}. This results in an overall dominant CMB-SNIa tension that renders the $\nloc$ model non-concordant and disfavors it with respect to $\lcdm$ by $\Delta \chi^2=5.9$, given \textit{BAPJ} data. Furthermore, a posteriori constraints from RSD data have also been studied in Ref. \cite{Dirian:2016puz} and they increase the overall tension even more, up to $\left. \Delta \chi^2 \right|_{\rm bapj} + \left. \Delta \chi^2 \right|_{\rm rsd}^{\rm post} = 8.5$ ($\sim 3 \sigma$) compared to $\lcdm$, ruling out the nonlocal model given \textit{BAPJ+{\rm(post)}RSD} data.

This problem can be solved by considering extensions of the initial model, that is, not only changing the dark energy parametrization as in Ref. \cite{Dirian:2016puz}, but also allowing other physically relevant parameters, otherwise fixed, to vary. Adding such new components will open new possibilities in the global parameter space and possibly provide an access to a new global maximum of the posterior probability distribution. The consequence of such a procedure is however the introduction of new degeneracies in the extended cosmological model coming together with a loss of constraining power for fixed data combination.

As already mentioned here-above, in this work we allow the sum of the neutrino masses to vary. This is justified for three reasons. 
First, constraints on $\sum m_\nu$ coming from terrestrial experiments are very weak, therefore there no obvious reason to fix $\sum m_\nu=0.06 \, {\rm eV}$ on empirical grounds. Indeed, this value only corresponds to the smallest mass-splitting measured by oscillations experiments (see subsection \ref{subsec:nu} for more details and references). 
Second, an increase in neutrino masses would not alter the expansion rate at early times if the neutrinos are still relativistic at photon decoupling, so that CMB anisotropies remain unaffected, and raise the energy density of pressureless matter (after they turn non-relativistic), which therefore increases the expansion rate during the matter dominated era. 
Third, the free-streaming behavior exhibited by a more massive neutrino component helps to tame the growth rate of structures and therefore potentially lowers the additional discrepancy caused by the inclusion of RSD data (see e.g. Refs. \cite{Archidiacono:2016lnv, Vagnozzi:2017ovm} for analyses of the effects induced by massive neutrinos on cosmological observables).
Thus, we can expect that the $RR$ model would prefer a higher value of $\sum m_\nu$ than $0.06 \, {\rm eV}$, with a corresponding decrease of the CMB-SNIa tension. This is what is discussed in more details in the following. 

The cyan curves in Fig.~\ref{fig:h} illustrate such facts in showing the same as the blue curves, but instead of adjusting $H_0$ to give the same $\theta_*$ as in $\lcdm$, one adjusts the neutrino masses to $\sum m_\nu = 0.42\ {\rm eV}$. The middle panel of Fig.~\ref{fig:h} also confirms that this would help to drastically improve the goodness-of-fit to \textit{Planck} data compared to the $RR$ model when using the $\lcdm$ best-fitting parameters. This is a clear sign that the degenerate effects of $H_0$, $\sum m_\nu$ and a late-time phantom dark energy on $\theta_*$ can therefore be exploited to try to reconcile the \textit{Planck} and SNIa \textit{JLA} constraints given the $RR$ nonlocal gravity.

Figure \ref{fig:hom} illustrates better the beneficial impact of varying $\sum m_\nu$ in the constraints given the $\nloc$ model compared to the $\lcdm$ one. The blue contours show the \textit{Planck} constraints on the $H_0 \!$ -- $ \! \Omega_{m}$ plane when $\sum m_\nu$ is a free parameter (quoted \textit{Planck} (free $\sum m_\nu$) for definiteness). For the nonlocal gravity model, the \textit{Planck} contour is now overlapping the SNIa one. This illustrates the fact that allowing $\sum m_\nu$ to vary weakens the CMB-SNIa tension, as quantified by the corresponding individual $\Delta \chi^2$ values reported in Table \ref{table:bapj}. 

A remarkable aspect of the combination of the \textit{Planck}, SNIa and BAO data in the constraints given $\nloc$ nonlocal gravity is the evidence for non-vanishing neutrino masses. Figure \ref{fig:tri} shows that $\sum m_\nu > 0$ at $\sim 2\sigma$ level, with the best-fit value $\sum m_\nu \approx 0.21\ {\rm eV}$. As depicted above, such a shift is primarily caused by the relatively smooth, late-time and phantom nature featured by the effective dark energy described by the $RR$ nonlocal model. These constraints are very different than in $\nu \lcdm$ for which the data only sets an upper bound on $\sum m_\nu$. In the following, we will see that such a preference of the $\nu \lcdm$ model for lower values of $\sum m_\nu$ reflects one of its weakness in a Bayesian model comparison context, and therefore opens room for alternative dark energy models including similarly a varying $\sum m_\nu$ to compete with it.

\subsection{Bayesian model comparison}

In the following, we compare the $\nu \lcdm$ and $\nu \nloc$ models given \textit{BAPJ} data computing the associated Bayes factor $B_{\nu \Lambda, \nu \nloc}$ and set it side by side with the Bayesian Information Criterion (BIC)\footnote{The Bayesian Information Criterion is given by (see e.g. Ref. \cite{Trotta_BITS} for more details),
\bq
{\rm BIC} \equiv \chi^2 + k \ln N \, , 
\eq
where $k$ is the number of parameters and $N$ the number of data points. The lower the BIC the better the model. Since $k$ and $N$ are equal within the models that we compare throughout this work, we can therefore only use the difference between the models $\chi^2$ as a BIC diagnostic. The latter criterion originates from an approximation of the Bayesian evidence assuming gaussianity of the posterior, a likelihood dominated regime and weak correlations between parameters.  The $\chi^2$ goodness-of-fit values computed throughout this work are obtained constructing low-temperature MCMC as described in more details in Sec. 2.2 of Ref.~\cite{Dirian:2016puz}. \label{ft:ft4}} 
differences that are reported in Table \ref{table:bapj}. Degrees of significance used in this work are reported in Table ~\ref{table:stat} for definiteness of the discussion. A Bayes factor $B_{01}$ comparing model $\mathcal{M}_0$ against model $\mathcal{M}_1$ can be though of as telling betting odds of $B_{01} \,$:$\, 1$ in favor of the former given the data.

For computing $B_{\nu \Lambda, \nu \nloc}$ we use a combination of statistical coherence and the Savage-Dickey density ratio (SDDR) (see Ref.~\cite{Dirian:2016puz} and references therein for details) that exploits the nested structure of the overall models discussed in that work. This allows one to get $B_{\nu \Lambda, \nu \nloc}$ in a rather economic way in writing,
\bq
B_{\nu \Lambda, \nu \nloc} &\equiv& \frac{P(d|\mathcal{M}_{\nu \Lambda})}{P(d|\mathcal{M}_{\nu \nloc})} \\
 &=& \frac{P(d|\mathcal{M}_{\nu \Lambda})}{P(d|\mathcal{M}_{\Lambda})} \frac{P(d|\mathcal{M}_{\Lambda})}{P(d|\mathcal{M}_{\nloc})} \frac{P(d|\mathcal{M}_{\nloc})}{P(d|\mathcal{M}_{\nu \nloc})} \\
&=& \frac{B_{\nloc,\nu \nloc}}{B_{\Lambda,\nu \Lambda} }  B_{\Lambda, \nloc} \, , \label{eq:bf_nLnRR}
\eq
where $P(d|\mathcal{M}_{i})$ is the marginal likelihood (evidence) of the data $d$ given the model $\mathcal{M}_{i}$. The factor $B_{\Lambda, \nloc}$ appearing above is one of the main results of Ref. \cite{Dirian:2016puz} and has been computed to be $B_{\Lambda, \nloc}=22.7$. The remaining factors are computed for the model $i=\Lambda,\nloc$ through the SDDR,
\bq
B_{i,\nu i} = \frac{P\big(\sum m_\nu=0.06 \, \big| \, d, \mathcal{M}_{\nu i}\big)}{P \big( \sum m_\nu=0.06 \, \big| \, \mathcal{M}_{\nu i} \big)} \, , \label{eq:sddr}
\eq
which is the ratio of the marginalized one-dimensional posterior distribution to the marginalized one-dimensional prior of $\sum m_\nu$ obtained from the extended model, evaluated at the point where the simpler model is nested inside the extended model (i.e. at $\sum m_\nu=0.06$).
Since we chose the same prior for both $\nu \lcdm$ and $\nu \nloc$ parameter spaces (in particular on $\sum m_\nu$) their contribution simplifies.  Eq.~\eqref{eq:bf_nLnRR} then yields,
\bq
B_{\nu \Lambda, \nu \nloc} &=& \frac{P\big(\sum m_\nu=0.06 \, \big| \, d, \mathcal{M}_{\nu \nloc}\big)}{P \big( \sum m_\nu=0.06 \, \big| \, d, \mathcal{M}_{\nu \Lambda} \big)} \, B_{\Lambda, \nloc} \, ,
\eq
and we find, given \textit{BAPJ} data, 
\bq
B_{\nu \Lambda, \nu \nloc} &=& \frac{1}{12.5} \times 22.7 = 1.8 = e^{0.6}\, ,
\eq
which is one of the main results of this article. It tells that $\nu \nloc$ is statistically equivalent to $\nu \lcdm$ with odds of $1.8 \,$:$\, 1$ for the latter, instead of being ``moderately-to-strongly'' disfavored with odds $22.7 \,$:$\, 1$ when the neutrino mass is fixed. The result is invariant under prior changes on $\sum m_\nu$ (as long as they are assumed to be equal) and leads to several implications. This result shows that allowing $\sum m_\nu$ to vary within $[0,5] {\rm eV}$ helps to reconcile $RR$-gravity with the data as already noticed above, but it also has the effect of penalizing the $\lcdm$ cosmology. This can be seen through the fact that applying the BIC method to compare $\nu \lcdm$ against $\nu \nloc$ given \textit{BAPJ} leads to biased results. Indeed, comparing the shifts endured by the results of both methods in varying $\sum m_\nu$ we obtain,
\begin{align}
 {\rm BIC:}& \left. \Delta \chi^2 \right|_{\Lambda, \nloc} \! \! \! &=&~~ 5.9  \! \! \!  &\rightarrow& \left. \Delta \chi^2 \right|_{\nu \Lambda, \nu \nloc} \! \! \! &=&~~ 3.4 \,,  \label{eq:chi2_evo}  \\ 
 {\rm Bayes:}& \left. \ln B_{ \Lambda, \nloc} \right. \! \! \! &=&~~ 3.1  \! \! \!  &\rightarrow& \left. \ln B_{\nu \Lambda, \nu \nloc} \right. \! \! \!  &=&~~ 0.6 \,, \label{eq:bf_evo}
\end{align} 
all in favor of $\lcdm$ given \textit{BAPJ} data. Referring to Table \ref{table:stat}, one can see a significant discrepancy between the results from the BIC differences and Bayes factors. While the former announces a reduction from ``weak''/``moderate-to-strong'' only to ``weak'' evidence in favor of $\nu \lcdm$, Bayesian model comparison tells that the latter ``moderate-to-strong'' evidence is in fact comfortably reduced to an ``inconclusive'' one. These two results are discrepant because of the loss of validity of the assumptions made in computing the BIC, which is only an approximation of the Bayes evidence (see footnote \ref{ft:ft4}). In particular as, beside of the net maximum likelihood shift encapsulated in $\Delta \chi^2$ favoring the $\nloc$ model, Occam's razor further penalizes the $\nu \lcdm$ one. Obviously one should therefore trust the result of Eq.~\eqref{eq:bf_evo}.

In what follows, we address a rough analysis for trying to understand to which extent allowing the absolute neutrino mass to vary is beneficial for the $\nloc$ model, given \textit{BAPJ} data. 
Coming back to Eq.~\eqref{eq:sddr}, in our present context we can write,
\bq
B_{i,\nu i} = \frac{P\big(\sum m_\nu=0.06 \, \big| \, d, \mathcal{M}_{\nu i}\big)}{P \big( \sum m_\nu=0.06 \, \big| \, \mathcal{M}_{\nu i} \big)} = \frac{L_i}{V_i} P_i \, ,
\eq
where $L_i$ is the value of the marginalized 1d posterior for $\sum m_\nu$ (normalized to its maximum) at the nesting point, $V_i$ is the volume of it and $P_i$ is the upper bound of the prior on the sum of the neutrino masses, $\sum m_\nu \in [0, P_i]$. We find,
\bq 
L_{\nloc} &=& 0.19 \quad, \qquad V_{\nloc}= 0.2 \, , \\
L_{\Lambda}&=& ~~ 1.0 \quad, \qquad V_{\Lambda} ~~\,= 0.08 \, ,
\eq
where we have set $P_i=5$ in either cases. One can then compute,
\bq 
B_{\Lambda, \nu \Lambda} = 12.5 \times 5 = 62.5 \quad, \, B_{\nloc, \nu \nloc} = 1 \times 5 \, ,
\eq
which in particular shows that the $\Lambda$-based model provides ``moderate-to-strong'' evidence with odds of $62.5 \,$:$\, 1$ for fixing $\sum m_\nu=0.06 {\rm ~ eV}$. Although the latter has a non-negligible contribution coming from the prior, it also has a non-negligible one from the likelihood (essentially originating from boundary effects), which is a handicap when compared against models preferring higher neutrino masses such as the nonlocal one studied in this work. As can be seen from Fig.~\ref{fig:tri}, this is because the $\nu \lcdm$ marginalized posterior on $\sum m_\nu$ hits the lower bound of the prior at $0.06 {\rm ~ eV}$, which involves a loss of posterior volume and a waste of prior one. In the $RR$ case the situation is different, since non-vanishing neutrino masses are preferred at $2 \sigma$ level, exploiting therefore better the \textit{BAPJ} data. This contributes to Occam's razor effect intrinsically taken into account in Bayesian model comparison and partially explains why the $\nloc$ nonlocal model undergoes a favorable and significant change when compared against $\lcdm$ after allowing $\sum m_\nu$ to vary (Eq.~\eqref{eq:bf_evo}). Moreover, this also explains why the BIC difference effectively fails when comparing $\nu \lcdm$ against $\nu \nloc$ given \textit{BAPJ} data, because it is only sensitive to the maximum of the posteriors, not to their entire volume.

The $\nloc$ nonlocal model described by the action \eqref{eq:action} is therefore statistically equivalent (given \textit{BAPJ} data) to Einstein gravity supplemented by a cosmological constant when reconsidering the prior on the neutrino sector, that is when one changes the cosmological parametrization from \eqref{base_param} to \eqref{nu_param}. This has been made possible exploiting an apparent degeneracy at the background level between $H_0$, $\sum m_\nu$ and the nature of the effective dark energy described by the nonlocal model, which was illustrated in Fig.~\ref{fig:hom}. In what follows we provide an outlook motivating the use of additional data, in particular coming from galaxy surveys, for being able to make a distinction between the $\nu \lcdm$ and the $\nu \nloc$ cosmological models. 

\begin{table}
\caption{Scale used for comparing model $\mathcal{M}_1$ against model $\mathcal{M}_0$ in this article, i.e. for interpetating their BIC difference $\Delta \chi^2_{01} \equiv \chi^2_1 - \chi^2_0$ and their log-Bayes factors $\ln B_{01}$. Positivity of the latter tends to favor $\mathcal{M}_0$. These scales are taken as a rule of thumb inspired by Secs.~2.6-2.10 of Ref.~\cite{opac-b1100695} in accordance with the (more conservative) Jeffreys' scale of Ref.~\cite{Trotta_BITS} (see also Ref.~\cite{Efstathiou:2008ed} for a comparison of the latter with the original scale proposed by Jeffrey).}
\begin{tabular}{@{}cccccccc}
\hline
\vspace{-0.25cm}\\
  &\ \ Interpretation \ \ &  $\Delta \chi^2$ \ \ &  $\ln B_{01}$ \ \ \\
  \vspace{-0.25cm}\\
\hline
\vspace{-0.25cm}\\
  &\ \ ``inconclusive'' \ \ &  $0 \,$--$\, 2$ \ \ &  $0 \,$--$\, 1$ \ \ \\
   &\ \ ``weak'' \ \ &  $2 \,$--$\, 6$ \ \ &  $1 \,$--$\, 2.5$ \ \ \\
   &\ \ ``moderate-to-strong'' \ \ &  $6 \,$--$\, 10$ \ \ &  $2.5 \,$--$\, 5$ \ \ \\
  &\ \ ``strong'' \ \ &  $>10$ \ \ & $>5$ \ \ \\
  \vspace{-0.25cm}\\
\hline
\end{tabular}
\label{table:stat}
\end{table}

\subsection{Constraints a posteriori from Redshift-Space Distortions data}

\begin{figure}
	\centering
	\includegraphics[scale=0.44]{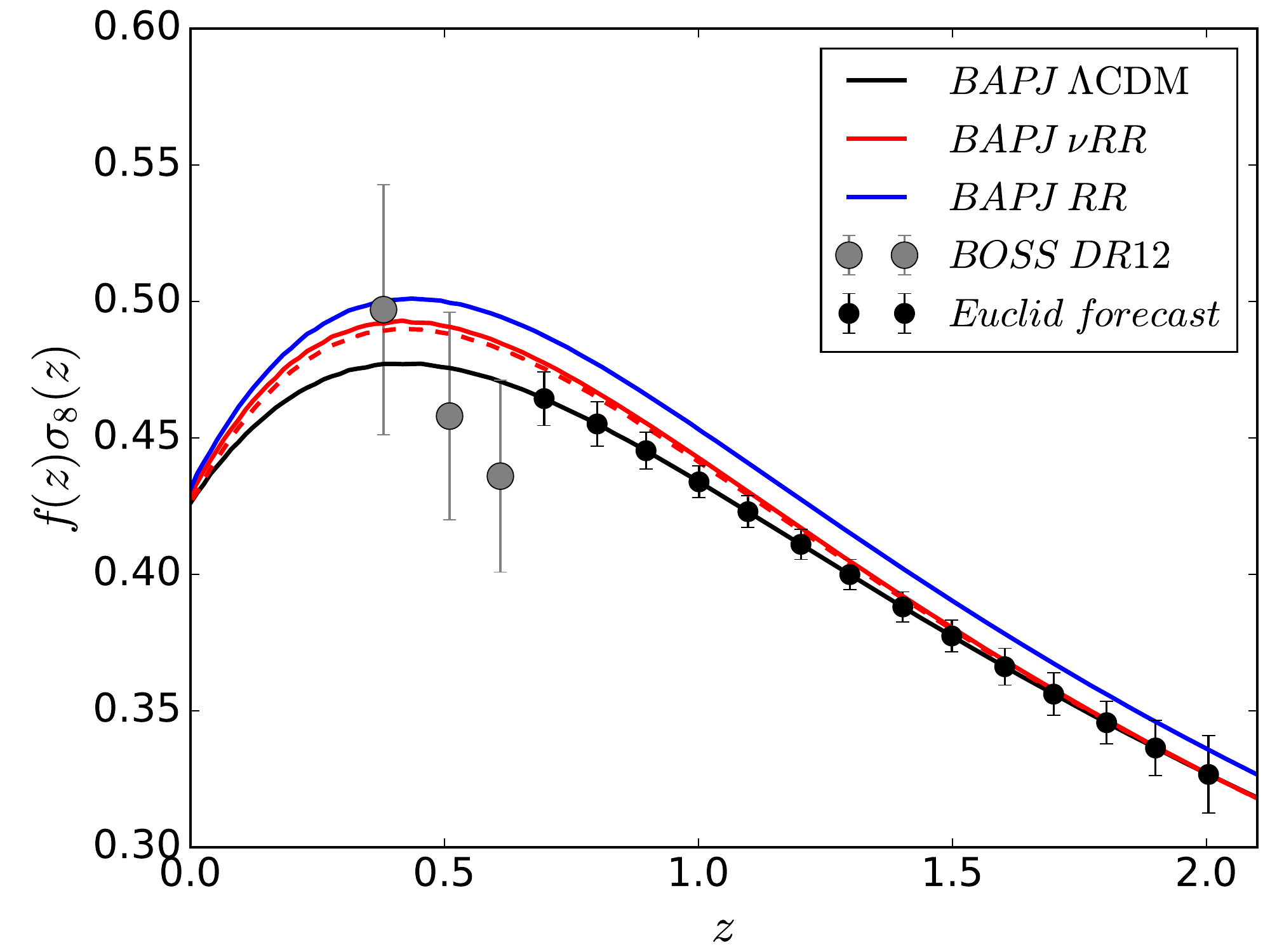}
	\caption{Time evolution of the growth rate $f\sigma_8$ for the best-fitting $\lcdm$, $RR$ and $\nu RR$ models to the \textit{BAPJ} dataset. For the case of the $\nu RR$ model (red), the solid and dashed lines display the result at $k = 0.01\ h/{\rm Mpc}$ and $k = 0.5\ h/{\rm Mpc}$, respectively. For the $\lcdm$ and $RR$ models the growth rate is scale-independent (apart from the very small scale-dependency induced by the small neutrino fraction, $\sum m_\nu = 0.06\ {\rm eV}$). The grey symbols show the observational determination from the final BOSS DR12 release \cite{Alam:2016hwk}.  The black symbols show the forecasted precision for Euclid, centered around the $\lcdm$ result.}
\label{fig:fs8}
\end{figure}

Apart from secondary CMB anisotropies such as ISW or lensing effects, the constraints on $\sum m_\nu$ mostly come from background-geometrical features when considering the \textit{BAPJ} dataset. However, massive neutrinos give rise to characteristic inhomogeneous and anisotropic signatures induced by their thermal velocity flow. In particular, they do not cluster inside regions delimited by their free-streaming scale. Below that scale, the neutrino perturbations are smoothed out and this causes a suppression of the late-time matter power spectrum at mid-to-small cosmological scales, a decrease of the lensing power and of the growth of structures in a scale-dependent manner within the linear regime \cite{2014NJPh...16f5002L, Hu:1997mj}, as well as non-linear effects \cite{Adamek:2016zes,Inman:2016qmg}. Additional data putting stronger constraints on these features are therefore relevant to include into the global fit. Nevertheless, the presence of an appreciable fraction of massive neutrinos can have partial degenerate effects with a positive fifth force present in modified gravity scenarios. A fifth force should be present at late-times in the $\nloc$ model and it was found in Refs.~\cite{2014JCAP...09..031B, Dirian:2016puz} that it enhances the growth of linear and non-linear structures compared to the one described by $\lcdm$. Into the latter reference, constraints on the linear growth rate of structure modelled by $f \sigma_8$ were derived given RSD data. It was found that the $\lcdm$ model was favored over the nonlocal one and this was quantified by a BIC of $\Delta \chi^2 = 2.6$. This value was computed a posteriori, that is, given that $f \sigma_8$ was derived from each model on its \textit{BAPJ} bestfit. In this part, we study the impact of a massive neutrino component on the linear growth rate of structure using the same method.

The degenerate effects present between a massive neutrino fraction and linear growth rate are well-illustrated from the degeneracy direction observed in the $\sigma_8 \,$--$\, \sum m_\nu$ plane in Fig.~\ref{fig:tri}, where one can see that they are anti-correlated: the higher the massive neutrino fraction $\Omega_\nu \sim \sum m_\nu$, the lower $\sigma_8$. Given \textit{Planck} data, the mean value inferred on $\sigma_8$ for $\lcdm$ is smaller than the one provided by $\nloc$, in agreement with the higher growth within the nonlocal model, and their mean values are generically smaller in the $\nu$-extended case. For \textit{Planck} only, we find that the departure of the best-fit value of $\sum m_\nu$ from the lower bound of the prior in $\nloc$ cosmology is caused by the addition of the \textit{Planck} CMB lensing power spectrum which is sensitive to a weighted projection of density fluctuations along the line-of-sight. Joining BAO+SNIa data pulls the total matter density fraction $\Omega_{m}$ to higher values, involving a stronger increase in the absolute neutrino mass that preserves the value of $\sigma_8$ close to the one inferred in $\lcdm$.
Focusing on the growth, Fig.~\ref{fig:fs8} shows the time evolution of $f\sigma_8$ for the best-fitting $\lcdm$, $RR$ and $\nu RR$ models to the \textit{BAPJ} dataset. As anticipated, the growth rate is lower in $\nu RR$ compared to the $RR$ model. The figure also displays the most recent observational determinations of $f\sigma_8$ from the DR12 BOSS analysis \cite{Alam:2016hwk} (grey symbols with errorbars). Using these data, the reduced $\chi_{\rm red}^2$ values for the $\lcdm$, $RR$ and $\nu RR$ models are, respectively\footnote{These values do not consider the mid-redshift data point. This is because the associated galaxy sample completely overlaps with those of the other two points which are independent. The number of degrees of freedom is therefore two.}, $\chi^2_{\rm red} = 0.58$, $\chi^2_{\rm red} = 1.38$ and $\chi^2_{\rm red} = 0.97$. More pragmatically for comparison with previous results, we compute the corresponding $\chi^2$ values using the same data points as in Ref. \cite{Dirian:2016puz}, that is, the ones collected from 6dF GRS \cite{Beu_6dF_RSD_2012} at $f \sigma_8(0.067)=0.423\pm0.055$, SDSS LRG \cite{Oka_SDSS_2013} at $f \sigma_8(0.3)=0.49\pm0.08$ , SDSS MGS \cite{How_SDSS_2014} at $f \sigma_8(0.15)=0.63^{+0.027}_{-0.24}$, BOSS LOWZ \cite{Chuang:2013wga} at $f \sigma_8(0.32)=0.371\pm0.091$, BOSS CMASS \cite{Samushia_BOSS_2013} at $f \sigma_8(0.57)=0.441\pm0.0434$ \footnote{Replacing the latter BOSS data by those of Ref. \cite{Alam:2016hwk} does not significantly affect our statistical conclusions.}, WiggleZ \cite{Blake_WiggZ_2012} at $f \sigma_8(0.44)=0.413\pm0.08$, $f \sigma_8(0.6)=0.39\pm0.063$, $f \sigma_8(0.73)=0.437\pm0.072$) and VIPERS \cite{delaTorre_VIPERS_2013} at $f \sigma_8(0.8)=0.47\pm0.08$. The corresponding goodness-of-fit read
\begin{align}
\chi^2_{\lcdm} = 3.9 \, , \qquad \chi^2_{\nloc} = 6.5 \, , \qquad \chi^2_{\nu \nloc} = 5.2 \, ,
\end{align}
which shows that the fit is indeed improved in going from $\nloc$ to $\nu \nloc$ with BIC values changing from $\Delta \chi^2 = 2.6$ to $\Delta \chi^2 = 1.3$ in favor of $\lcdm$. Therefore, we can conclude that allowing $\sum m_\nu$ to be a free parameter helps to decrease the discrepancy of the $\nloc$ nonlocal gravity model with growth rate measurements and brings down the total discrepancy from $\left. \Delta \chi^2 \right|_{\rm bapj} + \left. \Delta \chi^2 \right|_{\rm rsd}^{\rm post} = 8.5$ ($\sim 3 \sigma$) to $\left. \Delta \chi^2 \right|_{\rm bapj} + \left. \Delta \chi^2 \right|_{\rm rsd}^{\rm post} = 4.6$ ($\sim 2 \sigma$) given \textit{BAPJ+{\rm(post)}{\rm RSD}}, which induces a significant change in the (although approximated) statistical conclusion.

In turn, this shows that the data considered in this work do not possess enough constraining power to clearly distinguish between $\lcdm$ and $\nloc$ cosmologies. Nevertheless, the situation is expected to be different for a survey like \textit{Euclid} \cite{Amendola:2012ys}. This is illustrated by the black symbols in Fig.~\ref{fig:fs8}, which show an estimate of the forecast errorbars for this future mission (taken from Fig.~3 of Ref.~\cite{2012MNRAS.424.1392M}), centred around the $\lcdm$ prediction. One notes that the difference between $\lcdm$ and $\nu RR$ is larger than the forecast precision of \textit{Euclid} for $z < 1$, from which we can conclude that, despite partial degeneracies between the effects of massive neutrinos and the $\nloc$-modifications to gravity, there is still room for future RSD data to be used to help distinguishing between $\lcdm$ and $RR$ cosmologies. In addition, constraints using weak gravitational lensing data such as those of CFHTLenS \cite{2012MNRAS.427..146H} or KiDS \cite{Kuijken:2015vca} also prove to be of particular interest for constraining the $\nu$-extended models. Indeed, measuring the cosmic shear induced by the large-scale structure, these data allow to put constraints on the nature of the dark energy as well as on the absolute neutrino mass \cite{Joudaki:2016kym}, although to a smaller extent than the \textit{BAPJ} dataset considered in this work. However, weak lensing measurements reported in Refs. \cite{2012MNRAS.427..146H, Kuijken:2015vca} are in tension with \textit{Planck} CMB observations, given the $\lcdm$ model, and systematic issues first need to be addressed before these data can be used in combination with \textit{Planck} for constraining modified gravity models.

As a final remark, a larger fraction of massive neutrinos in cosmological models contributes to an increased scale-dependence in the linear growth of structure. This may raise some concerns when confronting models like the best-fitting $\nu RR$ model to \textit{BAPJ} against $f\sigma_8$ values, because the latter are usually extracted from galaxy survey data using RSD models assuming the growth to be scale-independent (see e.g.~Ref.~\cite{2014MNRAS.444.3926J} for an exception to this fact and Ref.~\cite{Barreira:2016ovx} for a validation study of RSD modelling in DGP gravity which exhibits scale-independent linear growth). If the scale-dependence in the $\nu RR$ is non-negligible compared to the precision targeted, extra care is required in the analysis of the data before observational constraints can be performed. To test such a fact, we plot in Fig.~\ref{fig:fs8} the evolution of $f\sigma_8$ in the $\nu RR$ model for $k = 0.01\ h/{\rm Mpc}$ (red solid) and $k = 0.5\ h/{\rm Mpc}$ (red dashed). One notes that the $k$-dependence is small compared to the expected precision of \textit{Euclid}, which suggests that standard methods can be used to constrain the $\nu RR$ model.

\subsection{A word on $H_0$}

Another interesting outcome of the constraints on the $RR$ model relates to the preferred values of $H_0$. For the best-fitting $\lcdm$ model to the \textit{BAPJ} dataset, one finds $H_0 = 67.67^{+0.47}_{-0.50}\ {\rm km/s/Mpc}$ (cf.~Table \ref{table:planck}), which lies $\sim 1\sigma$ below the determination from local measurements discussed in Ref.~\cite{2014MNRAS.440.1138E}, which sets $H_0 = 70.6 \pm 3.3\ {\rm km/s/Mpc}$ (note that this value becomes $H_0 = 72.5 \pm 2.5\ {\rm km/s/Mpc}$ if other assumptions are made into the analysis). More recently, the work of Ref.~\cite{2016ApJ...826...56R} sets a higher value $H_0 =  73.24 \pm 1.74\ {\rm km/s/Mpc}$ (see also Refs.~\cite{2011ApJ...730..119R, 2013ApJ...775...13H}). Furthermore, recent determinations of $H_0$ using hyperparameters, $H_0 = 73.75 \pm 2.11 \ {\rm km/s/Mpc}$ \cite{Cardona:2016ems}, or from gravitational lensing time delay methods, $H_0 = 71.9^{+2.4}_{-3.0} \ {\rm km/s/Mpc}$ \cite{2016arXiv160701790B}, are also significantly away from the $\lcdm$ bestfit.

The seriousness of the above-mentioned $H_0$ tensions is still subject to current debates and one still needs to understand better the role of systematics before claiming the need of new physics (see e.g.~Refs.~\cite{DiValentino:2016hlg, Lukovic:2016ldd, Bernal:2016gxb}). Nevertheless, taking the current measurements at face value, one notes that for the best-fitting $\nu RR$ model to the \textit{BAPJ} dataset one has $H_0 = 69.49^{+0.79}_{-0.80} \ {\rm km/s/Mpc}$, which significantly ameliorates the agreement with the local determinations and would therefore improve further the global fit. 

\subsection{The importance of terrestrial determinations of $\sum m_\nu$} \label{subsec:nu}

The current constraints on neutrino masses that are independent of cosmology arise from terrestrial experiments. The lower bounds on $\sum m_\nu$ come from neutrino oscillations experiments which, assuming a massless eigenstate, set $\sum m_\nu \gtrsim 0.05\ {\rm eV}$ and $\sum m_\nu \gtrsim 0.1\ {\rm eV}$ for normal and inverted mass hierarchies respectively. The current best upper bounds are obtained by analysing the high-energy part of the spectrum of Tritium $\beta$-decay in experiments such as MAINZ and TROITSK and set the electron neutrino mass to $m_{\nu_e} \lesssim 2.2 \ {\rm eV}$ ($2 \sigma$) which corresponds to $\sum m_\nu \lesssim 6.6 \ {\rm eV}$ in our context. Future Tritium $\beta$-decay experiments such as KATRIN will be sensitive to mass scales $\sum m_\nu \lesssim 0.6\ {\rm eV}$ at $90 \%$ confidence level. The sensitivity can be even better if neutrinos turn out to be Majorana particles, in which case neutrinoless double $\beta$ decay experiments should be able to probe the region corresponding to $\sum m_\nu \gtrsim  0.3\ {\rm eV}$ with high precision\footnote{Note that the quoted upper bounds assume specific models of nuclear matrix elements.} (see e.g.~Refs.~\cite{Drexlin:2013lha, DellOro:2016tmg, Vergados:2016hso, Engel:2016xgb} for reviews). These forecast sensitivities can therefore be proven useful for confirming cosmological observations. As it has been shown throughout our study, the determination of the absolute neutrino mass scale from cosmological probes depends on the assumed cosmological model.
As such, if terrestrial neutrino experiments will detect non-minimal neutrino masses, we will need to modify the standard $\lcdm$ cosmological model.

\section{Summary \& Conclusion}\label{sec:conc}

We have revisited the observational constraints of Ref. \cite{Dirian:2016puz} where the nonlocal model of modified gravity described in Sec.~\ref{sec:model} was found to be disfavored against $\lcdm$ with odds of $22:1$, given \textit{Planck}+SNIa+BAO data. Such a discrepancy was noticed to be mostly caused by a CMB-SNIa tension in the $\Omega_m \,$--$\, H_0$ plane, that we analyzed in more details into our Sec.~\ref{sec:res}. We have found that it results from the quite smooth, late-time and phantom nature of the effective dark energy described by the nonlocal model which, for fixed parameter values, induces a decrease on the late-time Hubble expansion rate $H(z \approx 0)$ compared to that described by $\lcdm$. Such a fact generically implies a smaller acoustic scale $\theta_*$ for the CMB which is corrected by the inference of a higher value of $H_0$ given \textit{Planck} data, as well as a larger luminosity distance that is compensated by a larger $\Omega_m$ given SNIa data. Since the shape information from CMB temperature power spectrum constrains well $\omega_m$, which is a multiplicative combination of $H_0$ and $\Omega_m$, the trends inferred from the nonlocal model are contradictory and a tension appears, as illustrated in Fig.~\ref{fig:hom}. 

We have then shown that allowing the absolute neutrino mass to be a free parameter in the nonlocal gravity model resolved the tension. The $\nu$-extended nonlocal model, denoted $\nu \nloc$, ends up to be statistically equivalent to $\nu \lcdm$ given \textit{Planck}+SNIa+BAO data with odds of $1.8:1$ in favor of the latter. We have shown that the compatibility between $\nu \nloc$ and $\nu \lcdm$ was caused by a better fit of the nonlocal model to the data, but also by the Occam's razor effect penalizing the $\nu \lcdm$ model because of its preference for small absolute neutrino masses.
As a result, the absolute neutrino mass is inferred to be non-zero $\sum m_\nu > 0$ at $\sim 2\sigma$ level given the nonlocal model, with the best-fitting value $\sum m_\nu \approx 0.21\ {\rm eV}$.
We have then placed constraints from RSD data a posteriori on both models, i.e. considering the matter power spectra corresponding to their respective \textit{Planck}+SNIa+BAO bestfit. These constraints have been shown to be improved as well by the presence of a higher neutrino fraction $\Omega_\nu$ into the nonlocal cosmology. Further determinations from local measurements of $H_0$ were also discussed, as these are in better agreement with the nonlocal gravity model inferring a value of $H_0 = 69.49^{+0.79}_{-0.80}\ {\rm km/s/Mpc}$, which is $\sim 2 \sigma$ above $H_0 = 67.67^{+0.47}_{-0.50}\ {\rm km/s/Mpc}$, inferred from $\lcdm$.

\medskip

In conclusion, letting the absolute neutrino mass to be a free parameter allowed the resulting $\nu \nloc$ nonlocal gravity model to fit current \textit{Planck}+SNIa+BAO data as well as $\nu \lcdm$. Given these data, the nonlocal model provides $\sum m_\nu > 0$ at $\sim 2\sigma$ with bestfit  $\sum m_\nu \approx 0.21\ {\rm eV}$. This is in disagreement with the value inferred from $\nu \lcdm$ which prefers $\sum m_\nu = 0\ {\rm eV}$, corresponding to the lower extreme value of the prior chosen in this work. Furthermore, allowing $\sum m_\nu$ to take higher values within the nonlocal cosmology also improved the fit to RSD data done a posteriori. This provides one more example showing that the cosmological constraints on the absolute neutrino mass depends on the assumed cosmological model, because of degenerate effects between modifications to gravity and massive neutrinos. Still, our study also suggests that the use of additional data coming from future galaxy redshift surveys could reduce such a degeneracy in the studied case and potentially discriminate between the $\nu \lcdm$ and $\nu \nloc$ models. We have provided an illustration to this fact in considering forecast constraints from \textit{Euclid} RSD data. A more quantitative analysis is left for future work. 

\begin{acknowledgments}

The author sincerely thanks Alexandre Barreira for his useful contributions to early stages of this work and for fruitful discussions. Many thanks to Martin Kunz and Michele Maggiore for their useful comments about the manuscript, to Le\"ila Haegel for sharing her expertise in experimental neutrino research and to the anonymous referee for many valuable comments contributing to the clarity of the manuscript. Numerical work presented in this publication used the Baobab cluster of the University of Geneva. YD is supported by the Fonds National Suisse.

\end{acknowledgments}

\bibliography{nonlocalnu.bib}

\end{document}